\documentstyle[aps,prb,array,multicol,psfig]{revtex}
\def\wide#1#2{
\end{multicols}
\widetext
\noindent
\if#1t
\else
    \raisebox{9pt}[0in][0.0in]
    {$\rule{3.4in}{0.4pt}\rule{0.4pt}{6pt}$\hspace{3.6in}}
\fi
#2
\if#1b
\else
    \raisebox{-9pt}[0in][0.0in]
    {\hspace{3.55in}$\rule{0.4pt}{6pt}\rule[6pt]{3.5in}{0.4pt}$}
\fi
\begin{multicols}{2}
\noindent
}

\begin{document}
\input{psfig.sty}

\title{Magnetic Quantization of Electronic States in $d$-wave Superconductors}

\author{N. B. Kopnin $^{(1,2)}$ and V. M. Vinokur $^{(3)}$}
\address{$^{(1)}$ Low Temperature Laboratory, Helsinki University of Technology,\\
P.O. Box 2200, FIN-02015 HUT, Finland,\\
$^{(2)}$ L. D. Landau Institute for Theoretical Physics, 117940 Moscow,
Russia,\\
$^{\left( 3\right) }$Argonne National laboratory, Argonne, Illinois 60439}
\title{Magnetic quantization of electronic states in $d$--wave superconductors}
\date{\today}
\maketitle

\begin{abstract}

We derive a general quasiclassical approach for long--range magnetic--field
quantization effects in superconductors. The method is applied to superclean 
$d$-wave superconductors in the mixed state. We study the delocalized states
with energies $\epsilon \gg \Delta _{0}\sqrt{H/H_{c2}}$. We find that the
energy spectrum consists of narrow energy bands whose centers are located at
the Landau levels calculated in absence of the vortex potential. We show
that transitions between the states belonging to the different Landau levels
give rise to resonances in the a.c. quasiparticle conductivity and in the
a.c.  vortex friction.
\end{abstract}


\begin{multicols}{2}

\section{Introduction}
The unusual behavior of thermodynamic and transport
properties of $d$--wave superconductors as
functions of magnetic field is being a subject of extensive
experimental and theoretical studies. This behavior is attributed to 
nontrivial energy dependence of the electronic density of states 
\cite{Volovik,VolKop-dos,SimonLee,STM}, and to specific kinetic 
processes which are
very sensitive to fine details of electronic states brought about by the 
presence of vortices \cite{Lee,Ong,Taillefer,KopVol-vdyn}.
  There exists, however, a conceptual controversy about the
structure of electronic states in $d$--wave superconductors in the mixed
state. One of the views is that the states below the maximum gap $\Delta _{0}$
have a discrete spectrum due to Andreev reflections; some states are
localized within vortex cores \cite{KopVol-vdyn,Kop/dwave} while others are
quantized at longer distances \cite{GorkSchrieff,Anderson,Janko} as a
particle moving along a curved trajectory in a magnetic field hits the gap
for a current momentum direction such that $\epsilon =\Delta _{{\bf p}}$.
Other authors advocate that, instead of the magnetic quantization, energy bands
should appear in a periodic vortex potential due to the vortex lattice 
\cite{Nielsen,Melnikov,Franz,Kita}.

In the present paper, we develop a general quasiclassical approach for
calculating the long--range magnetic--field quantization effects in
superconductors in the regime where the wave-length of electrons is 
much shorter
than the coherence length $p_{F}\xi \gg 1$. The proposed method is applied to 
superclean $d$--wave superconductors in the mixed state in the low field
limit, $H\ll H_{c2}$.  
We demonstrate that quantization effects are in fact a compromise between the 
two abovementioned extremes. In the first part of the paper (Sections 
\ref{LONGRANGE} - \ref{PERIODIC}),
we show that the influence of a magnetic field on delocalized
excitations in a superconductor cannot be reduced to simply an action of an
effective vortex lattice potential. The effect of magnetic field is rather 
two-fold: (i) It creates vortices and thus provides a periodic
potential for electronic excitations. (ii) It also affects the long range
motion of quasiparticles in a manner similar to that in the normal state. 
The latter long range effects are less pronounced for
low energy excitations. 
The spectrum of excitations with energies 
$\epsilon \gtrsim \Delta _{0}\sqrt{H/H_{c2}}$, however,
is mostly determined by the long range motion and exhibits magnetic 
quantization.

We study the delocalized states with energies $%
\epsilon \gg \Delta _{0}\sqrt{H/H_{c2}}$ and calculate their energy
spectrum.  We find that the spectrum indeed consists of energy bands
as it should be in a periodic potential.
However, in the quasiclassical limit, the bands are rather narrow;
their centers are located at the Landau levels calculated in Refs. \onlinecite
{GorkSchrieff,Anderson,Janko}.

In the second part, Sections \ref{TRANSITIONS} - 
\ref{CONDUCTIVITY}, we consider effects of the energy
spectrum on the vortex dynamics and on the quasiparticle conductivity. We
show that both the vortex friction for oscillating vortices and the a.c.
quasiparticle conductivity for fixed (pinned) vortices display resonances at
transitions between the states belonging to different Landau levels.

\section{Long--range effects of the magnetic field}\label{LONGRANGE}
We start with the conventional Bogoliubov--de Gennes equations 
\begin{eqnarray}
\left[ \left( \hat{{\bf p}}-\frac{e}{c}{\bf A}\right) ^{2}-p_{F}^{2}\right] {%
u}+2m\Delta _{\hat{{\bf p}}}{v} &=&2m\epsilon {u} \, , \nonumber \\
\left[ \left( \hat{{\bf p}}+\frac{e}{c}{\bf A}\right) ^{2}-p_{F}^{2}\right] {%
v}-2m\Delta _{\hat{{\bf p}}}^{*}{u} &=&-2m\epsilon {v}  \label{eq-bog}
\end{eqnarray}
where $\hat{{\bf p}}=-i\nabla $ is the canonical momentum operator.
Equations (\ref{eq-bog}) have the particle--hole symmetry such that $%
u\rightarrow v^{*},v\rightarrow -u$ under complex conjugation and $\epsilon
\rightarrow -\epsilon $.
For a vortex array, the order--parameter phase is a multiple-valued function
defined through 
\[
{\rm curl}\nabla \chi =\sum_{i}2\pi \delta ({\bf r}-{\bf r}_{i}) \, .
\]
As a result 
\begin{equation}
\nabla \chi =\sum_{i}\frac{{\bf z}\times ({\bf r}-{\bf r}_{i})}{|{\bf r}-%
{\bf r}_{i}|^{2}}  \label{phase}
\end{equation}
such that, on average, $\nabla \chi \approx eHr/c$ for large $r$.

Consider a quasiparticle in a magnetic field in the presence of the 
vortex lattice
for energies ranging from the above the gap to infinity. If the particle 
mean free path is
longer than the Larmor radius, i.e., $\omega _{c}\tau \gg 1$ where 
$\omega_{c}$ is the cyclotron frequency, such particle can travel away 
from each
vortex up to distances of the order of the Larmor radius 
$r_{L}=v_{F}/\omega_{c}$. This brings new features to Eqs. (\ref{eq-bog}). 
Assume for a
moment that $\Delta =0$. The wave function $u$ describes then a particle 
with the
kinetic momentum ${\bf P}_{+}={\bf p}-(e/c){\bf A}$ and an energy $\epsilon =%
{\bf P}_{+}^{2}/2m-E_{F}$ while $v$ describes a hole with the kinetic
momentum ${\bf P}_{-}={\bf p}+(e/c){\bf A}$ and an energy $\epsilon =E_{F}-%
{\bf P}_{+}^{2}/2m$. A particle and a hole which start propagation from the 
same point
will then move in different directions and along different trajectories
which transform one into another under the transformation ${\bf H}%
\rightarrow -{\bf H}$. For a finite order parameter, the wave function is a
linear combination of a particle and a hole. It is not convenient, however,
to use such a combination at distances where the trajectories of a particle 
and a hole go far apart, i.e., when the vector potential is no longer small
compared to the Fermi momentum $p_{F}$.

Eq. (\ref{eq-bog}) shows that the phase of $u$ differs from that of $v$ by
the order parameter phase $\chi$. To construct a proper basis, one needs to
bring the phases of $u$ and $v$ in correspondence with each other. We note
that the usual transformation 
\[\left( 
\begin{array}{c}
u \\ 
v
\end{array}
\right) =\left( 
\begin{array}{c}
e^{i\chi /2}\tilde{u} \\ 
e^{-i\chi /2}\tilde{v}
\end{array}
\right) 
\]
is not convenient when considering a particle which can move at distances much
larger than the size of one unit cell. The problem is that the new functions 
$\tilde{u}$ and $\tilde{v}$ have extra phase factors $\pm \chi /2$ as
compared to the initial functions $u$ and $v$, respectively. These phases
increase with the distance resulting thus in a shift in the action $%
A\rightarrow \tilde{A}\pm \chi /2$.  The latter and is equivalent to a 
shift in the
momentum ${\bf p}=\nabla A\rightarrow \tilde{{\bf p}}\pm \nabla \chi /2$. 
This transformation
is not dangerous if the particle is bound to distances of the order of one
intervortex distance because the phase gradient is limited $\left| \nabla
\chi \right| \ll p_{F}$. However, for a vortex array, the phase gradient
increases with distance and can reach values comparable with $p_{F}$. It
means that components of the new momentum can not be integrals of motion
(i.e., they change along the trajectory) even in absence of the vortex
potential associated with the superconducting velocity and spatial
variations of the order parameter magnitude.

To avoid these complications we use another transformation which also
removes the coordinate dependence of the order parameter phase. The results,
of course, should be independent of the choice of the transformation due to
the gauge invariance. Following Refs. \onlinecite{Anderson} and 
\onlinecite{Melnikov} we put in Eq. (\ref{eq-bog}) 
\begin{equation}
u=\tilde{u},\;v=\exp \left( -i\chi \right) \tilde{v} \, . \label{transf3}
\end{equation}
This is a single-valued transformation. We obtain 
\begin{eqnarray}
\left[ \hat{{\bf P}}_+ ^{2}-p_F^{2}\right] \tilde{u}+2me^{-i\chi }\Delta _{%
\hat{{\bf P}}_+^\prime }\tilde{v} =2m\epsilon \tilde{u}  \label{transf1}\, , \\
\left[ \left( \hat{{\bf P}}_+-2m{\bf v}_{s}\right) ^{2}-p_F^{2}\right] 
\tilde{v}-2me^{i\chi }\Delta _{\hat{{\bf P}}_+^\prime} ^{*}\tilde{u}
=-2m\epsilon \tilde{v}  \label{transf2}
\end{eqnarray}
where $\hat {{\bf P}}_+=\hat{{\bf p}}-\frac{e}{c}{\bf A}$ is the operator of
the particle kinetic momentum, and 
\[
\hat{{\bf P}}_+^\prime =\hat{{\bf p}}-\nabla \chi /2 =\hat{{\bf P}}_+ -m{\bf %
v}_{s}\, . 
\]
The superconducting velocity is 
\[
2m{\bf v}_{s}={\bf \nabla }\chi -\frac{2e}{c}{\bf A} \, .
\]
In Eqs. (\ref{transf1}), (\ref{transf2}) we use that, for a general pairing
symmetry, $\Delta _{\hat{{\bf p}}^\prime}\propto uv^*$ depends actually on $%
\hat{{\bf p}}^\prime =(\hat{{\bf p}}_{u}+\hat{{\bf p}}_{v})/2$ where $\hat{%
{\bf p}}_{u,v}$ are the canonical momentum operators which act on the
Bogoliubov wave functions $u$ and $v$, respectively. The term $-\nabla \chi
/2$ appears in the order parameter together with the canonical momentum $%
{\bf p}$ because only one half of the momentum operator in $\Delta _{\hat{%
{\bf p}}^\prime}$ acts on each of the wave functions $u$ or $v$.

The transformation of Eq. (\ref{transf3}) is ``$u$--like'' and brings the phase
of $v$ in correspondence with the phase of $u$. The resulting equations are
not symmetric with respect to $u$ and $v$: the term ${\bf v}_{s}$ is present
in the second equation together with $\hat{{\bf P}}$ while it does not
appear in the first equation. Let us perform one more transformation 
\begin{equation}
\left( 
\begin{array}{c}
\tilde{u} \\ 
\tilde{v}
\end{array}
\right) =\left( 
\begin{array}{c}
U \\ 
V
\end{array}
\right) e^{i\chi _{v}/2}  \label{transf4}
\end{equation}
where ${\bf \nabla }\chi _{v}=2m{\bf v}_{s}$ such that 
\[
{\rm curl}{\bf \nabla }\chi _{v}=\sum_{i}2\pi \delta ({\bf r}-{\bf r}_{i})-%
\frac{2e}{c}{\bf H}
\]
and $\chi _{v}=\chi -\chi _{A}$ where 
\begin{equation}
\chi _{A}=\frac{2e}{c}\int_{{\bf r}_{0}}^{{\bf r}}[{\bf H}\times {\bf r}%
^{\prime }]\,d{\bf r}^{\prime } \, . \label{chiA}
\end{equation}
The ``phase'' $\chi _{v}$ is not single valued within each unit cell, it
depends on the particular path of integration. However, it is single valued
on average, i.e., on a scale much larger than the intervortex distance since 
\[
\int {\rm curl}{\bf \nabla }\chi _{v}d^{2}r=0 \, .
\]
It also implies that $\chi _{v}$ does not have large terms increasing with
distance. The transformation Eq. (\ref{transf4}) is thus not dangerous. The
total transformation Eqs. (\ref{transf3},\ref{transf4}) has the form 
\begin{eqnarray}
u &=&\exp \left( i\chi /2-i\chi _{A}/2\right) U \, ,  \nonumber \\
v &=&\exp \left( -i\chi /2-i\chi _{A}/2\right) V \, . \label{transf6}
\end{eqnarray}
With this transformation we finally obtain 
\begin{eqnarray}
\left[ \left( {\bf \hat{P}}_{+}-m{\bf v}_{s}\right) ^{2}-p_{F}^{2}\right]
U+2m\tilde{\Delta}_{{\bf \hat{P}}_{+}}V &=&2m\epsilon U \, , \nonumber \\
\left[ \left( {\bf \hat{P}}_{+}+m{\bf v}_{s}\right) ^{2}-p_{F}^{2}\right]
V-2m\tilde{\Delta}_{{\bf \hat{P}}_{+}}U &=&-2m\epsilon V  \label{eq-bog2}
\end{eqnarray}
where 
\[
\tilde{\Delta}_{{\bf \hat{P}}_{+}}=e^{-i\chi }\Delta _{\hat{{\bf p}}-\left(
e/c\right) {\bf A}}=e^{i\chi }\Delta _{\hat{{\bf p}}-\left( e/c\right) {\bf A%
}}^{*} \, .
\]
As distinct from Eq. (\ref{eq-bog}), a particle and a hole determined by Eq.
(\ref{eq-bog2}) move along the same trajectory though, of course, in
different directions.

One can transform these equations further by putting 
\begin{equation}
\check \Psi = \left( 
\begin{array}{c}
U \\ 
V
\end{array}
\right) =\exp \left( i\int {\bf p}\cdot d{\bf r}\right) \check \phi ~;~ 
\check \phi = \left( 
\begin{array}{c}
\phi _1 \\ 
\phi _2
\end{array}
\right)  \label{transf-quasi}
\end{equation}
where 
\begin{equation}
\left( {\bf p}-\frac{e}{c}{\bf A}\right) ^{2}=p_{F}^{2} \, . \label{p-magn}
\end{equation}
If ${\rm div}{\bf A}=0$ we have 
\begin{eqnarray}
{\bf P}_{+}\left( -i\nabla +m{\bf v}_{s}\right) \phi _1 +m\tilde{\Delta}_{%
{\bf P}_{+}}\phi _2 &=&m\epsilon \phi _1  \, ,\nonumber \\
{\bf P}_{+}\left( -i\nabla -m{\bf v}_{s}\right) \phi _2 -m\tilde{\Delta}_{%
{\bf P}_{+}}\phi _1 &=&-m\epsilon \phi _2 \, . \label{eq/qcl}
\end{eqnarray}

Another equation can be obtained using the transformation 
\begin{eqnarray}
u &=&e^{i\chi }e^{-i\chi _{v}/2}U=\exp \left( i\chi /2+i\chi _{A}/2\right)
U\, ,  \nonumber \\
v &=&e^{-i\chi _{v}/2}V=\exp \left( -i\chi /2+i\chi _{A}/2\right) V \, .
\label{transf5}
\end{eqnarray}

We get 
\begin{eqnarray}
\left[ \left( {\bf \hat{P}}_{-}-m{\bf v}_{s}\right) ^{2}-p_{F}^{2}\right]
U+2m\tilde{\Delta}_{{\bf \hat{P}}_{-}}V &=&2m\epsilon U \, , \nonumber \\
\left[ \left( {\bf \hat{P}}_{-}+m{\bf v}_{s}\right) ^{2}-p_{F}^{2}\right]
V-2m\tilde{\Delta}_{{\bf \hat{P}}_{-}}U &=&-2m\epsilon V  \label{eq-bog2*}
\end{eqnarray}
where $\hat{{\bf P}}_{-}=\hat{{\bf p}}+\left( e/c\right) {\bf A}$ is the
``hole'' kinetic momentum. The transformation Eq. (\ref{transf5}) is ``$v$%
--like'', it brings the phase of $u$ in correspondence with that of $v$.
Using Eq. (\ref{transf-quasi}) we can transform Eq. (\ref{eq-bog2*}) to its
quasiclassical version which is Eq. (\ref{eq/qcl}) where ${\bf P}_+$ is
substituted with ${\bf P}_-$ under the condition $|{\bf P}_-|^2=p_F^2$. Eq. (%
\ref{eq/qcl}) and its $v$-like analogue possess the particle--hole symmetry.
Under transformation 
\[
{\bf p}\rightarrow -{\bf p},\epsilon \rightarrow -\epsilon ;\;\phi _1
\rightarrow \phi _2^{*},\phi _2\rightarrow -\phi _1^{*} 
\]
they go one into another. Moreover, each set of equations has the
particle--hole symmetry separately for a given position on the trajectory if
the kinetic momenta ${\bf P}_{\pm }={\bf p}\mp \left( e/c\right) {\bf A}$
are reversed for a fixed position of the particle. Due to Eq. (\ref{p-magn}) 
${\bf p}-(e/c){\bf A}=\left( q\cos \alpha ,\,q\sin \alpha \right) $, where $%
\alpha $ is the local direction of the momentum. The reversal corresponds to 
$\alpha \rightarrow \pi +\alpha $.

We take the $z$ axis along the magnetic field and define the quasiclassical
particle-like trajectory in Eq. (\ref{eq/qcl}) by 
\begin{equation}
\frac{dx}{dy}=\frac{p_{x}-(e/c)A_{x}}{p_{y}-(e/c)A_{y}}\, .  \label{traject}
\end{equation}
When the magnetic filed penetration length is much longer than the distance
between vortices, $\lambda _{L}\gg a_{0}$, the magnetic field can be
considered homogeneous. With ${\bf A}$ taken in the Landau gauge 
\begin{equation}
{\bf A}=\left( -Hy,\,0,\,0\right)   \label{Landau-gauge}
\end{equation}
the trajectory is a circle 
\begin{equation}
\left( x-x_{0}\right) ^{2}+\left( y+cp_{x}/eH\right) ^{2}=\left( p_{\perp
}c/eH\right) ^{2}  \label{Landtraject}
\end{equation}
where $p_{\perp }^{2}=p_{F}^{2}-p_{z}^{2}$. The local direction of the
kinetic momentum is $p_{x}+eHy/c=p_{\perp }\sin \alpha $, $p_{y}=p_{\perp
}\cos \alpha $. The distance along the trajectory is $ds=r_{L}d\alpha $
where the Larmor radius is $r_{L}=p_{\perp }/m\omega _{c}$.

Eq. (\ref{eq/qcl}) has a simple physical meaning. It is the quasiclassical
version of the usual Bogoliubov--de Gennes equation for vortex state
modified to take into account long range effects of magnetic field. Eq. (\ref
{eq/qcl}) can be written in terms of the particle trajectory Eq. (\ref
{traject}). We have from Eq. (\ref{eq/qcl}) 
\begin{eqnarray}
v_{\perp }\left( -i\frac{\partial }{\partial s}+mv_{t}\right) \phi _{1}+%
\tilde{\Delta}\left( \alpha \right) \phi _{2} &=&\epsilon \phi _{1} 
\, ,\nonumber \\
v_{\perp }\left( -i\frac{\partial }{\partial s}-mv_{t}\right) \phi _{2}-%
\tilde{\Delta}\left( \alpha \right) \phi _{1} &=&-\epsilon \phi _{2}
\, .\label{eq/traject}
\end{eqnarray}
Here $v_{\perp }=p_{\perp }/m$, and $v_{t}$ is the projection of ${\bf v}_{s}
$ on the local direction of the trajectory. $\Delta (\alpha )$ and $v_{t}$
are functions of coordinates $x\left( s\right) $, $y\left( s\right) $, and
of the angle $\alpha (s)$ taken at the trajectory. Eqs. (\ref{eq/traject})
look exactly as the usual Bogoliubov--de Gennes equations.

\section{Electronic states in zero lattice potential}\label{ZEROPOTENTIAL}

For $d$--wave superconductors, we take the order parameter in the form $%
\tilde{\Delta }_{{\bf p}}=\Delta _{0}\left( 2p_{x}p_{y}\right) /\left(
p_{x}^{2}+p_{y}^{2}\right) $ so that $\tilde{\Delta }_{{\bf p}-(e/c){\bf A}%
}=\Delta _{0}\sin (2\alpha ) $. Consider first the limit $v_{s}=0$ and $%
\Delta _0 =const$. Eqs. (\ref{eq/traject}) become 
\begin{eqnarray*}
-i\omega _{c}\frac{\partial \phi _1}{\partial \alpha } +\Delta _{0}\sin
(2\alpha )\phi _2 &=&\epsilon \phi _1 \, ,\\
i\omega _{c}\frac{\partial \phi _2}{\partial \alpha } +\Delta _{0}\sin
(2\alpha )\phi _1 &=&\epsilon \phi _2 \, .
\end{eqnarray*}
With 
\[
\check \phi =\check C \exp \left[ if\left( \alpha \right) \right] 
\]
we obtain 
\[
f\left( \alpha \right) =\pm \int \frac{d\alpha }{\omega _{c}}\sqrt{\epsilon
^{2}-\Delta _{0}^{2}\sin ^{2}(2\alpha )} \, .
\]

The quantization rule also includes the integral over the momentum ${\bf p}$
defined by Eqs. (\ref{transf-quasi}, \ref{p-magn}). We have 
\begin{equation}
\oint {\bf p}\,d{\bf r}\pm \oint \frac{d\alpha }{\omega _{c}}\sqrt{\epsilon
^{2}-\Delta _{0}^{2}\sin ^{2}(2\alpha )}=2\pi n  \, .\label{BS-rule}
\end{equation}

The quasiclassical approximation holds for $n\gg 1$. The $\pm $ signs
distinguish between particles and holes. As it was already mentioned, a
particle (with the plus sign in Eq. (\ref{BS-rule})) and a hole (with the
minus sign) move along the same trajectory but in the opposite directions.
The phase $\chi _{v}$ which was introduced in Eqs. (\ref{transf4}, \ref
{transf6}) gives a contribution to the action of the order of $2\pi $
because it is limited from above by an increment of the order of circulation
around one vortex unit cell; it can thus be neglected for large $n$.

\subsection{Sub-gap states}

In the range $|\epsilon |<\Delta _{0}$, the turning points correspond to 
vanishing of
the square root at $\alpha =\pm \alpha _{\epsilon }$ where $\sin (2\alpha
_{\epsilon })=|\epsilon |/\Delta _{0}$. We have 
\begin{equation}
\frac{4}{\omega _{c}}\int_{0}^{\alpha _{\epsilon }}d\alpha \sqrt{\epsilon
^{2}-\Delta _{0}^{2}\sin ^{2}(2\alpha )}=2\pi n  \label{quant-local}
\end{equation}
where $n>0$. The first integral in Eq. (\ref{BS-rule}) disappears because
the turning points of the momentum ${\bf p}$ are not reached: the particle
can not go far along the trajectory Eq. (\ref{traject}) and remains
localized on a given trajectory at distances $s\sim r_{L}(\epsilon /\Delta
_{0})$ smaller than the Larmor radius $r_{L}$. Note also that the
contribution from $\chi _{v}$ vanishes identically because the particle
after being Andreev reflected transforms into a hole which returns to the
starting point along the same trajectory. Using the substitution $\sin
x=(\Delta _{0}/\epsilon )\sin \left( 2\alpha \right) $ we find 
\begin{eqnarray*}
\int_{0}^{\alpha _{\epsilon }}d\alpha \sqrt{\epsilon ^{2}-\Delta
_{0}^{2}\sin ^{2}(2\alpha )} &=&\frac{\Delta _{0}}{2}\left[ E\left( \frac{%
\epsilon }{\Delta _{0}}\right) \right.  \\
&&-\left. \left( 1-\frac{\epsilon ^{2}}{\Delta _{0}^{2}}\right) K\left( 
\frac{\epsilon }{\Delta _{0}}\right) \right] 
\end{eqnarray*}
where $K\left( k\right) $ and $E\left( k\right) $ are the full elliptic
integrals of the first and second kind, respectively. Applying the
Bohr--Sommerfeld quantization rule Eq. (\ref{BS-rule}) we obtain 
\begin{equation}
\frac{2\Delta _{0}}{\omega _{c}}\left[ E\left( \frac{\epsilon _{n}}{\Delta
_{0}}\right) -\left( 1-\frac{\epsilon _{n}^{2}}{\Delta _{0}^{2}}\right)
K\left( \frac{\epsilon _{n}}{\Delta _{0}}\right) \right] =2\pi n \, .
\label{E/local}
\end{equation}
These states are degenerate with the same degree as in the normal state: for
each $n$, there are $\Phi /2\Phi _{0}=N_{v}/2$ states for particles and $%
N_{v}/2$ states for holes, where $\Phi $ is the total magnetic flux through
the superconductor, and $N_{v}$ is the total number of vortices.

Consider $\epsilon \ll \Delta _{0}$. Expanding in small $k$%
\[
E\left( k\right) =\frac{\pi }{2}\left( 1-\frac{k^{2}}{4}\right) ,\;K\left(
k\right) =\frac{\pi }{2}\left( 1+\frac{k^{2}}{4}\right) 
\]
we find from Eq. (\ref{E/local}) 
\begin{equation}
\epsilon _{n}=\pm \sqrt{4\Delta _{0}\omega _{c}\,n} \, . \label{E-GS}
\end{equation}
Eq. (\ref{E-GS}) agrees with the result of Refs. 
\onlinecite{GorkSchrieff},\onlinecite{Anderson}.

\subsection{Extended states}

If $|\epsilon |>\Delta _{0}$, we get for the Landau gauge Eq. (\ref
{Landau-gauge}) $p_{x}=const$ and 
\begin{eqnarray*}
\oint {\bf p}\,d{\bf r}=\oint p_{y}\,dy &=&2\int_{y_{1}}^{y_{2}}\sqrt{%
p_\perp ^{2}-(p_{x}+eHy/c)^{2}}\,dy \\
&=&\pi cp_\perp ^{2}/eH \, .
\end{eqnarray*}
The turning points $y_{1,2}$ correspond to the values of Larmor radius where 
$p_{x}+eHy_{1,2}/c=\pm p_\perp $. The corresponding trajectory is a closed 
circle where $%
\alpha $ varies by $2\pi $. The second integral in Eq. (\ref{BS-rule}) gives 
\begin{equation}
\int_{0}^{2\pi }\frac{d\alpha }{\omega _{c}}\sqrt{\epsilon ^{2}-\Delta
_{0}^{2}\sin ^{2}(2\alpha )}=\frac{4\epsilon }{\omega _{c}}E\left( \frac{%
\Delta _{0}}{\epsilon }\right) \, . \label{pint2}
\end{equation}
The quantization rule (\ref{BS-rule}) yields 
\begin{equation}
\pm \frac{2\epsilon }{\pi }E\left( \frac{\Delta _{0}}{\epsilon _n}\right)
=\omega _{c}n+\frac{p_{z}^{2}}{2m}-E_{F} \, . \label{E-deloc}
\end{equation}
For an $s$--wave superconductor we get, in particular, 
\begin{equation}
\pm \sqrt{\epsilon _n^{2}-\Delta _{0}^{2}}=\omega _{c}n+\frac{p_{z}^{2}}{2m}%
-E_{F} \, . \label{E-swave}
\end{equation}

\section{Effects of the periodic potential}\label{PERIODIC}

\subsection{Bloch functions}

At low magnetic fields $H\ll H_{c2}$, one can consider that the particle
trajectory always passes  far from cores. The oscillating part of the order
parameter comes mostly from the superconducting velocity. The corresponding
Doppler energy $\eta =p_{\perp }v_{t}$ is of the order of $\Delta _{0}\sqrt{%
H/H_{c2}}$. This periodic potential can split the energy spectrum into bands.
Eqs. (\ref{eq-bog2}, \ref{eq-bog2*}) or the quasiclassical version Eq. (\ref
{eq/qcl}) are invariant under the magnetic translations by periods of the
regular vortex lattice.  Consider the particle-like equations (\ref{eq-bog2})
or (\ref{eq/qcl}).  The particle-like operator of magnetic translations in a
homogeneous field is \cite{Brown} 
\begin{equation}
\hat{T}\left( {\bf R}_{l}\right) =\exp \left[ -i{\bf R}_{l}\left( \hat{{\bf p%
}}+\frac{e}{c}{\bf A}\right) \right]   \label{magtransoper}
\end{equation}
where $\hat{{\bf p}}=-i\nabla $ is the canonical momentum and ${\bf R}_{l}$
is a vector of the vortex lattice. Its zero--field version corresponds to a
shift 
\[
\hat{T}_{0}\left( {\bf R}_{l}\right) f\left( {\bf r}\right) =\exp \left[ -i%
{\bf R}_{l}\hat{{\bf p}}\right] f\left( {\bf r}\right) =f\left( {\bf r}-{\bf %
R}_{l}\right) \, .
\]
The operator $\hat{T}\left( {\bf R}_{l}\right) $ commutes with the
Hamiltonian because ${\bf v}_{s}$ and $\Delta $ are periodic in the vortex
lattice and the commutator 
\[
\left[ \left( \hat{{\bf p}}+\frac{e}{c}{\bf A}\right) _{i},\,\left( \hat{%
{\bf p}}-\frac{e}{c}{\bf A}\right) _{j}\right] =0\, .
\]
Since ${\bf P}_{+}$ does not change under the action of the operator Eq. (%
\ref{magtransoper}), magnetic translations for functions $\check{\phi}$ in
Eq. (\ref{eq/qcl}) are equivalent to usual translations $\hat{T}_{0}({\bf R}%
_{l})$ in space for a fixed kinetic momentum of the particle.

It is more convenient to consider magnetic translations in the symmetric
gauge ${\bf A}={\bf H}\times {\bf r}/2$. In this case, 
\[
\hat{T}\left( {\bf R}_{l}\right) f\left( {\bf r}\right) =\exp \left( -\frac{%
ie}{2c}{\bf R}_{l}\left[ {\bf H}\times {\bf r}\right] \right) f\left( {\bf r}%
-{\bf R}_{l}\right) \, .
\]
For this gauge, the wave functions Eq. (\ref{transf-quasi}) can be more
conveniently written in a slightly different form 
\begin{equation}
\check{\Psi}\left( p_{x};\,{\bf r}\right) =\exp \left[
ieHxy/2c+ip_{x}x+i\int_{y_{1}}^{y}p_{y}\,dy^{\prime }\right] \check{\phi}\, .
\label{Landaufucn1}
\end{equation}
The extra phase factor $\exp [ieHxy/2c]$ is associated with our choice of
the vector potential and allows to reduce the problem to the Landau gauge.
The particle trajectory takes the form of Eq. (\ref{Landtraject}) with $%
p_{y}=\sqrt{p_{\perp }^{2}-\left( p_{x}+eHy/c\right) ^{2}}$. The function $%
\check{\phi}$ satisfies Eq. (\ref{eq/traject}).

If $a_{0}$ and $b_{0}$ are the unit cell vectors along $x$ and $y$,
respectively, the magnetic translation operators for functions of Eq. (\ref
{Landaufucn1}) are 
\begin{eqnarray}
\hat{T}_{x}(la_{0})\check{\Psi}_{n}(p_{x};\,x,y) &=&e^{-ip_{x}la_{0}}\check{%
\Psi}_{n}(p_{x};\,x,y)  \,  ,\label{magtrans3} \\
\hat{T}_{y}(lb_{0})\check{\Psi}_{n}(p_{x};\,x,y) &=&\check{\Psi}_{n}(p_{x}-%
\frac{eHlb_{0}}{c};\,x,y)\, .
\end{eqnarray}
When deriving these expressions we have used the periodicity of ${\bf v}_{s}$ 
and
the fact that the trajectory depends on $y$ only through $y+cp_{x}/eH$. The
turning point $y_{1}$ is thus shifted by $lb_{0}$ when $p_{x}$ is shifted by 
$-eHlb_{0}/c$.

The functions Eq. (\ref{Landaufucn1}) can be used to construct two
independent basis functions 
\begin{eqnarray}
\check{\Phi}_{n}^{+}\left( k_{x},k_{y};x,y\right) &=&
\sum_{l}e^{ik_{y}2lb_{0}}\hat{T}_{y}(2lb_{0})\check{\Psi}_{n}(k_{x};\,x,y)
\, , \label{Blochfunct} \\
\check{\Phi}_{n}^{-}\left( k_{x},k_{y};x,y\right) &=&
\sum_{l}e^{ik_{y}(2l+1)b_{0}}  \nonumber \\
&&\times \hat{T}_{y}\left((2l+1)b_{0}\right) \check{\Psi}_{n}(k_{x};\,x,y)
\label{Blochfunct-}
\end{eqnarray}
with even and odd translations, respectively. Starting from Eq. (\ref
{Blochfunct}) we replace $p_x$ with $k_x$. The functions $\check \Phi ^{\pm}$
belong to the same energy. The wave vector $k_y$ has an arbitrary value, we
shall establish it later. The generic translation is $2b_0$ which is the
size of the magnetic unit cell along the $y$ axis. The magnetic unit cell
contains two vortices because the superconducting magnetic flux quantum
correspond to one half of the $2\pi$ phase circulation of a single--particle
wave function. The functions Eqs. (\ref{Blochfunct}, \ref{Blochfunct-}) have
the Bloch form 
\begin{eqnarray}
\hat{T}_{x}(la_{0})\check{\Phi}^{\pm}(k_{x},k_{y}) &=& (\pm 1)^l
e^{-ik_{x}la_{0}}\check{\Phi}^{\pm}(k_{x},k_{y}) \, , \label{Bloch1} \\
\hat{T}_{y}(2mb_{0})\check{\Phi}^{\pm}(k_{x},k_y ) &=&e^{-ik_{y}2mb_{0}} 
\check{\Phi}^{\pm}(k_{x},k_{y}) \, . \label{Bloch2}
\end{eqnarray}
We omit the coordinates $x,y$ in the arguments of $\check{\Phi}^{\pm}$ for
brevity. The functions $\check \Phi ^{\pm}$ transform into each other under
odd translations 
\begin{equation}
\hat{T}_{y}\left((2m+1)b_{0}\right)\check{\Phi}^{\pm}(k_{x},k_y ) =
e^{-ik_{y}(2m+1)b_{0}} \check{\Phi}^{\mp}(k_{x},k_{y}) . \label{Bloch3}
\end{equation}

Since the magnetic translation $\hat{T}_{y}(lb_{0})$ commutes with the
Hamiltonian, the energy is degenerate with respect to $k_{y}$. This
degeneracy is spurious, however. To see this, consider the transformations
Eqs. (\ref{Bloch1}, \ref{Bloch2}). For $l=1$, the transformed function in
Eq. (\ref{Bloch1}) is periodic in $k_{x}$ with the period $2\pi /a_{0}$.
This period corresponds to the shift of the center of orbit $y_{0}=ck_{x}/eH$
by one size of the magnetic unit cell $2b_{0}$. Obviously, the
transformation Eq. (\ref{Bloch2}) should also have the same symmetry. For
one magnetic unit cell, a shift by $2b_{0}$ (i.e., for $m=1$) along the $y$
axis should combine with one period along the $x$ axis. The period in $k_{y}$
is $\pi /b_{0}$; it should thus correspond to the shift of the coordinate $%
x_{0}$ by $a_{0}$. This fixes 
\begin{equation}
k_{y}=eHx_{0}/c  \, .\label{x0}
\end{equation}
The energy depends on the position of the trajectory within the vortex unit
cell through the Doppler energy $\eta $. The energy $\epsilon (k_{x},k_{y})$
has a band structure due to periodicity of $\eta $; it is periodic with the
periods $eHb_{0}/c=\pi /a_{0}$ and $eHa_{0}/c=\pi /b_{0}$ in $k_{x}$ and $%
k_{y}$, respectively, which correspond to shifts of the center of orbit by
one vortex unit cell vector.

\subsection{Spectrum}

Consider energies $\epsilon \gg \Delta _{0}\sqrt{H/H_{c2}}$. Applying the
quasiclassical approximation to Eq. (\ref{eq/traject}) we find 
\begin{equation}
\check{\phi}=\check{C}\exp \left[ \pm iA(s)\right]  \label{phi}
\end{equation}
where the action is
\begin{equation}
A(s)=\int_{s_{1}}^{s}\sqrt{\left( \epsilon -\eta \right) ^{2}-\Delta
_{0}^{2}\sin ^{2}(2\alpha )}\frac{ds}{v_{\perp }} \, . \label{action}
\end{equation}
The quasiclassical approximation is justified because the wave vector $%
\partial A/\partial s\sim \epsilon /v_{F}$ is much larger than the inverse
characteristic scale $1/a_0$ of variation of the potential $\eta $ for $%
\epsilon \gg \Delta _{0}\sqrt{H/H_{c2}}$. The function $\eta =p_{\perp
}v_{t} $ is taken at the trajectory which is a part of a circle specified by
the coordinates of its center $x_{0}$ and $y_{0}=-cp_{x}/eH$; they determine
the position of the trajectory within the vortex unit cell.

For energies $\Delta _{0}\sqrt{H/H_{c2}}\ll \epsilon < \Delta _{0}$,
quasiparticle trajectory is extended over distances of the order of $%
r_{L}\left( \epsilon /\Delta _{0}\right) $. The quantization rule defines
the energy 
\begin{equation}
\int_{s_{1}}^{s_{2}}\sqrt{\left( \epsilon -\eta \right) ^{2}-\Delta
_{0}^{2}\sin ^{2}(2\alpha )}\frac{ds}{v_{\perp }}=\pi n  \, . \label{quant}
\end{equation}
Here $s_{1}$ and $s_{2}$ are the turning points. Expanding in small $\eta
\ll \epsilon $ we find 
\begin{eqnarray}
&&m\int_{y_{1}}^{y_{2}}\sqrt{\epsilon ^{2}-\Delta _{0}^{2}\sin ^{2}(2\alpha )%
}\frac{dy}{p_{y}}  \nonumber \\
&-&m\int_{y_{1}}^{y_{2}}\frac{\eta (x,y)\epsilon }{\sqrt{\epsilon
^{2}-\Delta _{0}^{2}\sin ^{2}(2\alpha )}}\frac{dy}{p_{y}}=\pi n
\, . \label{quant1}
\end{eqnarray}
Here $\eta (x,y)=(k_{x}+eHy/c)v_{sx}+p_{y}v_{sy}$ while $y_{1}$ and $y_{2}$
are the turning points which correspond to vanishing of the square root: $%
k_{x}+eHy_{1,2}/c=p_{\perp }\sin (2\alpha _{\epsilon })$. The energy $%
\epsilon _{n}$ is a function of $k_{x}$ and $x_0$ which determine the
location of the particle trajectory with respect to vortices. The energy is
thus periodic in $k_x$ with the period $eHb_0/c$ and in $x_0$ with a period $%
a_0$ when the center is shifted by one period of the vortex lattice.

The $\eta $ term under the second integral in Eq. (\ref{quant1}) oscillates
rapidly over the range of integration and mostly averages out. The remaining
contribution determines the variations of energy with $k_{x}$ and $x_{0}$
and can be estimated as follows. For example, variation of action for $%
\epsilon \ll \Delta _{0}$ due to a change in energy $\delta \epsilon $ is 
\[
\delta A\sim (\delta \epsilon /v_{F})(\epsilon /\Delta _{0})r_{L}\sim
(\epsilon \delta \epsilon )/(\Delta _{0}\omega _{c})\, .
\]

Variation of action due to a shift of the center of orbit by a distance of
the order of the lattice period is $\delta A\sim (a_{0}/v_{F})\eta \sim 1$.
The corresponding energy variation is thus $\delta \epsilon \sim \Delta
_{0}\omega _{c}/\epsilon $. Since $x_{0}$ is coupled to $k_{y}$ through Eq. (%
\ref{x0}) the energy can be written as 
\begin{equation}
\epsilon _{n}\left( k_{x},k_{y}\right) =\sqrt{4\Delta _{0}\omega _{c}\left[
n+\eta _{0}\left( k_{x},k_{y}\right) \right] }  \label{bands}
\end{equation}
where $\eta _{0}\sim 1$ can depend on energy. The energy Eq. (\ref{bands})
has a band structure; the bandwidth is of the order of the distance between
the Landau levels. It is small as compared to the energy itself. It is clear
that the spectrum for energies $\epsilon \gtrsim \Delta _{0}$ can also be
obtained from Eqs. (\ref{E/local}), (\ref{E-deloc}) and (\ref{E-swave})
through the substitution $n\rightarrow n+\eta _{0}\left( k_{x},k_{y}\right) $%
.

One can check that, for given $k_x$ and $k_y$, the quasiparticle states with
different principle quantum numbers indeed concentrate near the levels
determined by Eq. (\ref{E-GS}) if $\epsilon \gg \Delta _{0}\sqrt{H/H_{c2}}$.
This is because the contribution to the action from the oscillating
potential picked up on the distance of the order of the size of the unit
cell $a_{0}$ is $p_{F}v_{s}a_{0}/v_{F}\sim 1$. The discrete structure of the
levels Eq. (\ref{E-GS}) would be preserved if the contribution to the action
from the oscillating potential changes by an amount much less than unity for
transitions between the neighboring levels. For an energy $\epsilon $, the
distance between the neighboring levels is $\delta \epsilon \sim \Delta
_{0}\omega _{c}/\epsilon $. This corresponds to a change in the length of
the trajectory by 
\[
\delta s_{\epsilon }\sim \frac{v_{F}}{\omega _{c}}\frac{\delta \epsilon }{%
\Delta _{0}}\sim \frac{v_{F}}{\epsilon } \, .
\]
The variation in the length is much smaller that the intervortex distance $%
a_{0}\sim \xi \sqrt{H_{c2}/H}$ if $\epsilon \gg \Delta _{0}\sqrt{H/H_{c2}}$,
and the action changes by a quantity much less than 1. It shows that the
distance between the levels with different $n$ is indeed determined by Eq. (%
\ref{E-GS}).

The situation changes for smaller energies $\epsilon \lesssim \Delta _0 
\sqrt{H/H_{c2}}$: The centers of bands will deviate strongly from positions
determined by Eq. (\ref{E-GS}) due to a considerable contribution from the
periodic vortex potential to the turning points in Eq. (\ref{quant}).
Moreover, the applicability of the quasiclassical approximation, i.e., of
Eq. (\ref{quant}) itself is violated; the potential $\eta$ is strong enough
to cause large deformations of the energy spectrum. As was shown in Ref. 
\onlinecite{Kop/dwave} some states can even become effectively localized 
near the vortex cores.

\section{Induced transitions between the Landau levels}\label{TRANSITIONS}

Vortex motion induces transitions between the quasiparticle states. The
transitions between low-energy core states with $\epsilon \ll \Delta _{0}%
\sqrt{H/H_{c2}}$ were considered in Ref. \onlinecite{Kop/dwave}. 
It was shown that
the vortex core states determine the vortex response to d.c. and a.c.
electric fields. For temperatures $T_{c}\sqrt{H/H_{c2}}\ll T$ extended
states dominate. It was found in Ref. \onlinecite{Kop/dwave} that the vortex
response is determined by what was called ``collective modes'' which are
associated with the electron states outside the vortex cores. In this
Section we demonstrate that these collective modes are nothing but
transitions between the electronic states Eq. (\ref{bands}) specified by the
same quasimomentum but by different principal quantum numbers $n$. We start
with noting that the transition matrix elements are proportional to 
\cite{KS} $%
\left\langle \check{\Phi}_{n}\left( {\bf k}_{i}\right) \nabla \check{H}_{1}%
\check{\Phi}_{m}\left( {\bf k}_{j}\right) \right\rangle $ where the
Hamiltonian $\check{H}_{1}$ is composed of $\Delta _{{\bf P}}$ and $\eta $,
while ${\bf k}$ is the quasimomentum. $\check{H}_{1}$ is periodic with the
period of the vortex lattice thus the transitions are possible between the
quasimomenta which differ by vectors of the reciprocal lattice. Since the
band energy is periodic in the quasimomenta with the periods of the
reciprocal lattice, the energy difference for these transitions corresponds
to the energy difference for states with the same quasimomentum but with
different quantum numbers $n$. For $\eta _{0}\ll n$ the transition energy is
just the distance between the Landau levels: $\delta \epsilon
_{n}(k_{x},k_{y})=\delta \epsilon _{n}$ determined by Eqs. (\ref{E/local}, 
\ref{E-deloc}) or (\ref{E-swave}). For low energies in a $d$--wave
superconductor, one has $\delta \epsilon \left( k_{x},k_{y}\right) =2\Delta
_{0}\omega _{c}/\epsilon _{n}$ in accordance with Eq. (\ref{E-GS}).

Consider transitions between the levels which are excited by oscillating the
vortices in more detail. We use the microscopic kinetic-equation approach
which has been applied earlier for $s$--wave superconductors in Ref. 
\onlinecite{KL}. The kinetic equations for the distribution functions 
${\sl f}_1$ and ${\sl f}_2$ have the form \cite{KB} 
\wide{m}{
\begin{eqnarray}
\left[ e\left( {\bf v}_{F}\cdot {\bf E}\right) g_{-}+\frac{1}{2}\left( f_{-}%
\frac{\hat{\partial}\Delta _{{\bf p}}^{*}}{\partial t}+f_{-}^{\dagger }\frac{%
\hat{\partial}\Delta _{{\bf p}}}{\partial t}\right) \right] \frac{\partial 
{\sl f}^{(0)}}{\partial \epsilon }+\left( {\bf v}_{F}\cdot \nabla \right)
(g_{-}{\sl f}_{2})+g_{-}\frac{\partial {\sl f}_{1}}{\partial t} &&  \nonumber
\\
+\left[ \frac{e}{c}\left[ {\bf v}_{F}\times {\bf H}\right] g_{-}-\frac{1}{2}%
\left( f_{-}\hat{\nabla}\Delta _{{\bf p}}^{*}+f_{-}^{\dagger }\hat{\nabla}%
\Delta _{{\bf p}}\right) \right] \cdot \frac{\partial {\sl f}_{1}}{\partial 
{\bf p}}+\frac{1}{2}\left( f_{-}\frac{\partial \Delta _{{\bf p}}^{*}}{%
\partial {\bf p}}+f_{-}^{\dagger }\frac{\partial \Delta _{{\bf p}}}{\partial 
{\bf p}}\right) \cdot \nabla {\sl f}_{1} &=&J  \label{kineq1}
\end{eqnarray}
}
and 
\begin{equation}
g_{-}\left( {\bf v}_{F}\cdot \nabla \right) {\sl f}_{1}=0.  \label{kineq2}
\end{equation}
Here 
\[
\check g^{R(A)}=\left( 
\begin{array}{cc}
g^{R(A)} & f^{R(A)} \\ 
-f^{\dagger R(A)} & -g^{R(A)}
\end{array}
\right) 
\]
are the retarded (advanced) quasiclassical Green functions, and $\check g_-
= \left(\check g^R -\check g^A \right)/2 $.

For an extended state with an energy $\epsilon >\Delta _{{\bf p}}$, the
particle trajectory crosses many vortex unit cells at various distances from
vortices. Since the distribution function ${\sl f}_{1}$ is constant along
the trajectory according to Eq. (\ref{kineq2}), it should be also
independent of the impact parameter (i.e., of the distance from the
trajectory to the vortex). We thus look for a distribution function ${\sl f}%
_{1}$ which is independent of coordinates. One can then omit the last term
in the l.h.s. of Eq. (\ref{kineq1}). Let us average Eq. (\ref{kineq1}) over
an area which contains many vortex unit cells but has a size small compared
with the Larmor radius, $a_{0}\ll r\ll r_{L}$. Since $r\ll r_{L}$ the
momentum ${\bf p}$ is still an integral of motion. We have (compare with 
Ref. \onlinecite{KB}) 
\begin{eqnarray*}
\int_{S_{0}}g_{-}\frac{\partial {\sl f}_{1}}{\partial t}\,d^{2}r-\frac{1}{2%
}{\rm Tr}\int_{S_{0}}d^{2}r\,\check{g}_-\left( \nabla \check{H}\right) %
\cdot \frac{\partial {\sl f}_{1}}{\partial {\bf p}}%
-\int_{S_{0}}J\,d^{2}r \\
=\frac{1}{2}{\rm Tr}\int_{S_{0}}d^{2}r\,\check{g}_{-} \left( {\bf v}_{L}\cdot
\nabla \check{H}\right) \frac{\partial {\sl f}^{(0)}}{%
\partial \epsilon }\, .
\end{eqnarray*}
Here Tr is the trace in the Nambu space, $S_{0}=\Phi _{0}/B$ is the area of
the vortex unit cell, 
\begin{eqnarray*}
J &=&-\frac{1}{\tau }\left[ \left( {\sl f}_{1}\left\langle
g_{-}\right\rangle -\left\langle {\sl f}_{1}g_{-}\right\rangle \right)
g_{-}\right.  \\
&&-\left. \left( {\sl f}_{1}\left\langle f_{-}^{\dagger }\right\rangle
-\left\langle {\sl f}_{1}f_{-}^{\dagger }\right\rangle \right) f_{-}+\left( 
{\sl f}_{1}\left\langle f_{-}\right\rangle -\left\langle {\sl f}%
_{1}f_{-}\right\rangle \right) f_{-}^{\dagger }\right] .
\end{eqnarray*}
Using the identity 
\[
\frac{1}{2}{\rm Tr}\int_{S_{0}}d^{2}r\,\left[ \left( \nabla \check{H}\right) 
\check{g}_{-}\right] =\pi \left[ {\bf z}\times {\bf v}_{\perp }\right] 
\]
derived in Ref. \onlinecite{KL} we find 
\begin{eqnarray}
&&-\pi \left[ {\bf z}\times {\bf v}_{\perp }\right] \cdot \frac{\partial 
{\sl f}_{1}}{\partial {\bf p}}+\frac{\partial {\sl f}_{1}}{\partial t}%
\int_{S_{0}}g_{-}\,d^{2}r-\int_{S_{0}}J\,d^{2}r  \nonumber \\
&=&\pi \left( {\bf v}_{L}\cdot \left[ {\bf z}\times {\bf v}_{\perp }\right]
\right) \frac{\partial {\sl f}^{(0)}}{\partial \epsilon } \, . 
\label{averkineq}
\end{eqnarray}

We shall concentrate on energies $\epsilon \gg \Delta \sqrt{H/H_{c2}}$. In
the leading approximation 
\begin{eqnarray*}
g_{-}&=&\frac{\epsilon }{\sqrt{\epsilon ^{2} -\Delta ^{2}\left( \alpha
\right) }} \Theta \left[ \epsilon ^{2}-\Delta ^{2}\left( \alpha \right)
\right] \, , \\
f_{-}&=&\frac{\Delta \left( \alpha \right) } {\sqrt{\epsilon ^{2}-\Delta
^{2}\left( \alpha \right) }}\Theta \left[ \epsilon ^{2}-\Delta ^{2}\left(
\alpha \right) \right] \, .
\end{eqnarray*}
We have 
\[
\left\langle {\sl f}_{1}\right\rangle =\left\langle {\sl f}%
_{1}g_{-}\right\rangle =0;~ \left\langle {\sl f}_{1}f_{-}\right\rangle
=\left\langle {\sl f}_{1}f_{-}^{\dagger }\right\rangle =0 \, .
\]
For a $d$--wave superconductor also $\left\langle f_{-}\right\rangle
=\left\langle f_{-}^{\dagger }\right\rangle =0$.

In the collision integral, the main contribution for $\epsilon \gg \Delta 
\sqrt{H/H_{c2}}$ comes from the delocalized states. Indeed, including
contributions from the bound states in the core \cite{KL} with energies $%
E_{n}(b)$ we would have 
\[
\int_{S_{0}}J\,d^{2}r\approx -S_{0}\left[ \sum_{n}\frac{p_{\perp }\omega _{c}%
}{\tau _{n}}\int \delta \left( \epsilon -E_{n}\right) db+\frac{\left\langle
g_{-}\right\rangle g_{-}}{\tau }\right] {\sl f}_{1}
\]
where $b$ is the impact parameter. The first term in square brackets comes
from the core states. Since $\tau _{n}\sim \tau $ and $b\sim \xi \sqrt{%
H_{c2}/H}$, the core contribution is of the order of $\tau ^{-1}\sqrt{%
H/H_{c2}}$. The delocalized states, however, give $\left( \epsilon /\Delta
_{0}\right) \tau ^{-1}$ which is much larger than the first term. Neglecting
the core contribution we find 
\[
J=-\frac{1}{\tau }\left\langle g_{-}\right\rangle g_{-}{\sl f}_{1} \, .
\]

Let us put 
\begin{equation}
{\sl f}_{1}=-\frac{\partial {\sl f}^{(0)}}{\partial \epsilon} [([{\bf %
u\times p}_{\perp }]\cdot {\bf \hat{z}})\gamma _{{\rm O}}+({\bf u\cdot p}%
_{\perp })\gamma _{{\rm H}}]  \label{distrfunc}
\end{equation}
The functions $\gamma _{{\rm O,H}}$ satisfy the following set of equations 
\begin{eqnarray}
\frac{\partial \gamma _{{\rm O}}}{\partial \alpha }-\gamma _{{\rm H}%
}-V\left( \alpha \right) \gamma _{{\rm O}}+1 &=&0  \nonumber \\
\frac{\partial \gamma _{{\rm H}}}{\partial \alpha }+\gamma _{{\rm O}%
}-V\left( \alpha \right) \gamma _{{\rm H}} &=&0  \label{eq/gammas}
\end{eqnarray}
which is derived  from Eq. (\ref{averkineq}). Here 
\begin{equation}
V\left( \alpha \right) =\frac{\left( -i\omega +\left\langle
g_{-}\right\rangle /\tau \right) g_{-}}{\omega _{c}} \, .  \label{V-def}
\end{equation}

The general solution of Eqs. (\ref{eq/gammas}) can be obtained \cite
{KopVol-vdyn} by putting $W_{\pm }=\gamma _{{\rm H}}\pm i\gamma _{{\rm O}}$.
We have 
\[
\frac{\partial W_{\pm }}{\partial \alpha }\mp iW_{\pm }-V\left( \alpha
\right) W_\pm \pm i=0 \, .
\]
The solution is 
\begin{equation}
W_{\pm }=\left[ C_{\pm }\mp i\int_{0}^{\alpha }e^{\mp i\alpha ^{\prime
}-F\left( \alpha ^{\prime }\right) }d\alpha ^{\prime }\right] e^{\pm i\alpha
+F\left( \alpha \right) }  \label{Wsolution}
\end{equation}
where 
\[
F\left( \alpha \right) =\int_{0}^{\alpha }V\left( \alpha ^{\prime }\right)
d\alpha ^{\prime } \, .
\]
The constant $C_{\pm }$ is found from the condition of periodicity $W\left(
\alpha \right) =W\left( \alpha +\pi /2\right) $%
\begin{equation}
C_{\pm }=\frac{\exp \left[ F\left( \pi /2\right) \right] \int_{0}^{\pi
/2}\exp \left[ \mp i\alpha -F\left( \alpha \right) \right] d\alpha }{1-\exp
\left[ \pm i\pi /2+F\left( \pi /2\right) \right] } . \label{C}
\end{equation}

In the limit $\tau \rightarrow \infty $, the solution 
\[
\gamma _{{\rm H}}=\left( W_{+}+W_{-}\right) /2;\;\gamma _{{\rm O}}=\left(
W_{+}-W_{-}\right) /2i 
\]
has poles when 
\begin{equation}
F\left( \pi /2\right) =\frac{\pi i}{2}(1+2M)  \label{res-cond}
\end{equation}
where $M$ is an integer. If $\epsilon >\Delta _{0}$, one obtains resonances
at 
\[
\frac{\omega }{\omega _{c}}\int_{0}^{2\pi }\frac{\epsilon }{\sqrt{\epsilon
^{2}-\Delta ^{2}\left( \alpha \right) }}d\alpha =2\pi \left( 1+2M\right)\, . 
\]
The lowest frequency $M=0$ exactly corresponds to the condition 
\[
\omega =\left( d\epsilon _{n}/dn\right) 
\]
where $d\epsilon _{n}/dn$ is the distance between the Landau levels
determined by Eq. (\ref{BS-rule}).

Note that, for an $s$--wave superconductor, Eqs. (\ref{eq/gammas}) has the 
form 
\begin{eqnarray}
\gamma _{{\rm H}}+V\gamma _{{\rm O}} &=&1  \nonumber \\
\gamma _{{\rm O}}-V\gamma _{{\rm H}} &=&0  \label{kineq-swave}
\end{eqnarray}
where 
\[
V\left( \alpha \right) =\left[ \frac{-i\omega }{\omega _{c}}\frac{\epsilon }{%
\sqrt{\epsilon ^{2}-\Delta _{0}^{2}}}+\frac{1}{\omega _{c}\tau }\right]
\Theta \left[ \epsilon ^{2}-\Delta _{0}^{2}\right] 
\]
since 
\[
J=-\frac{1}{\tau }\left( {\sl f}_{1}-\left\langle {\sl f}_{1}\right\rangle
\right) \Theta \left[ \epsilon ^{2}-\Delta _{0}^{2}\right] \, .
\]
One has from Eq. (\ref{kineq-swave}) 
\[
\gamma _{{\rm H}}=\frac{1}{1+V^{2}};~\gamma _{{\rm O}}=\frac{V}{1+V^{2}}\, . 
\]
The resonances appear when $\omega _{c}\tau \gg 1$; the poles correspond to $%
V=\pm i$ so that 
\[
\omega =\omega _{c}\frac{\sqrt{\epsilon ^{2}-\Delta _{0}^{2}}}{\epsilon }=%
\frac{d\epsilon _{n}}{dn} 
\]
where $\epsilon _{n}$ is determined by Eq. (\ref{E-swave}).

\subsection{Low energies}

For energies $\epsilon <\Delta _{0}$, the resonance condition Eq. (\ref
{res-cond}) is not just the distance between the Landau levels determined by
Eq. (\ref{E/local}). One has from Eq. (\ref{quant-local}) 
\[
\frac{d\epsilon _{n}}{dn}\int_{-\alpha _{\epsilon }}^{\alpha _{\epsilon }} 
\frac{\epsilon}{\sqrt{\epsilon ^{2}-\Delta ^{2}\left( \alpha \right) }}
d\alpha =\pi \omega _{c} 
\]
where $\Delta \left( \alpha _{\epsilon }\right) =\epsilon$. At the same
time, Eq. (\ref{res-cond}) gives the lowest resonant frequency 
\[
\frac{\omega }{\omega _{c}}N\int_{-\alpha _{\epsilon }}^{\alpha _{\epsilon }}%
\frac{\epsilon}{\sqrt{\epsilon ^{2}-\Delta ^{2}\left( \alpha \right) }}
d\alpha =2\pi 
\]
where $N$ is the number of gap nodes ( $N=4$ for a $d$--wave
superconductor). We see that the resonance occurs at 
\begin{equation}
N\omega =2\frac{d\epsilon _{n}}{dn}  \, . \label{res-cond1}
\end{equation}
When the vortex oscillates, all $N$ nodes participate in exciting
quasiparticles which accounts for the factor $N$ in the l.h.s. of Eq. (\ref
{res-cond1}). This is similar to the process of multi-photon absorption. The
factor $2$ in the r.h.s. is explained by noting that states with momentum
directions $\alpha$ and $\alpha +\pi$ are simultaneously excited.

Solution of Eqs. (\ref{eq/gammas},\ref{V-def}) for $\Delta _{0}\sqrt{H/H_{c2}%
}\ll \epsilon \ll \Delta _{0}$ was obtained in Ref. \onlinecite{Kop/dwave}. For
the main region of angles, $\left| \alpha \right| >\alpha _{\epsilon
}=\epsilon /2\Delta _{0}$. According to Eqs. (\ref{Wsolution}, \ref{C}), it
is 
\begin{eqnarray}
\gamma _{{\rm O}} &=&A\cos \alpha +B\sin \alpha  \nonumber \\
\gamma _{{\rm H}} &=&1-A\sin \alpha +B\cos \alpha  \label{gammas}
\end{eqnarray}
with 
\begin{equation}
A=\frac{e^{\lambda }\sinh \lambda }{2\sinh ^{2}\lambda +1};\;B=\frac{%
e^{-\lambda }\sinh \lambda }{2\sinh ^{2}\lambda +1} \, . \label{constants}
\end{equation}
Here we use that $F\left( \pi /2-\alpha \right) =2\lambda -F\left( \alpha
\right) $ where 
\[
\lambda =F\left( \alpha _{\epsilon }\right) ;F\left( \pi /2\right) =2\lambda 
\, .
\]
One has 
\[
\lambda =\frac{-i\omega +1/\tau _{eff}}{\omega _{c}} \frac{\pi \left|
\epsilon \right| }{4\Delta _{0}} 
\]
where $1/\tau _{eff}=\left| \epsilon \right|/\Delta _{0}\tau $ since $%
\left\langle g_{-}\right\rangle =\left| \epsilon \right|/\Delta _{0}$. Note
that a $\tau $-approximation was used in Ref. \onlinecite{Kop/dwave} for the
collision integral. To get the present expression for $\lambda $ from that
obtained in Ref. \onlinecite{Kop/dwave} one has to replace 
$1/\tau $ with $1/\tau _{eff}$.
For $\tau \rightarrow \infty $, the response Eqs. (\ref{gammas}), (\ref
{constants}) has poles at $i\lambda =\left( 2M+1\right) \pi /4$, i.e., for 
\begin{equation}
\omega =(2M+1)E_{0}(\epsilon );~E_{0}(\epsilon )=\Delta _{0}\omega
_{c}/\left| \epsilon \right| \, . \label{resonance}
\end{equation}
We have for $M=0$%
\[
\omega =\frac{1}{2}\frac{d\epsilon _{n}}{dn} 
\]
where $\epsilon _{n}$ is determined by Eq. (\ref{E-GS}). This condition
agrees with Eq. (\ref{res-cond1}).

These resonances were first predicted in Ref. \onlinecite{Kop/dwave}. Note the
different numerical factor in Eq. (\ref{resonance}) as compared to Ref. 
\onlinecite{Kop/dwave}; this is because a simplified version of 
$V\left( \alpha \right)$ has been used in Ref. \onlinecite{Kop/dwave}. 
The main effect of resonances is
that vortices experience a considerable friction force Eq. (\ref{fricforce})
even in a superclean case $\omega \tau _{eff}\gg 1$.

\subsection{Vortex friction}

A vortex moving with a velocity ${\bf v}_L$ experiences a force from the
environment 
\begin{equation}
{\bf F}_{{\rm env}}=-D{\bf v}_L- D^\prime [{\bf v}_L\times {\bf z}]\, .
\label{fricforce}
\end{equation}
According to Ref. \onlinecite{KL}, the delocalized states contribute to the
friction constant 
\begin{equation}
D_{{\rm del}}=\pi N\left\langle \int _{{\rm del}}\gamma _{{\rm O}} \frac{%
df^{(0)}}{d\epsilon} \frac{d\epsilon}{2}\right\rangle _\alpha
\end{equation}
where $\left\langle \cdots\right\rangle _\alpha $ is an average over $d\alpha
$. The factor $D^\prime$ is determined by the same expression where $\gamma
_{{\rm O}}$ is replaced with $\gamma _{{\rm H}}$.

The presence of resonances makes the dissipative constant $D_{{\rm del}}$
finite even in the superclean limit $\omega _{c}\tau \rightarrow \infty $.
Indeed, for an $s$--wave case, 
\[
\gamma _{{\rm O}}=\frac{\pi E}{2}\left[ \delta (\omega -E)+\delta (\omega
+E)\right] 
\]
where $E=\omega _{c}\sqrt{1-\Delta _{0}^{2}/\epsilon ^{2}}$. The friction
constant becomes 
\begin{equation}
D_{{\rm del}}=\pi ^{2}N\Delta _{0}\frac{\omega ^{2}/\omega _{c}^{2}}{\left(
1-\omega ^{2}/\omega _{c}^{2}\right) ^{3/2}}\frac{df^{(0)}(\epsilon _{0})}{%
d\epsilon }
\end{equation}
where $\epsilon _{0}=\Delta _{0}/\sqrt{1-\omega ^{2}/\omega _{c}^{2}}$. A
more detailed discussion of the resonant vortex friction for a $d$--wave
superconductor at low temperatures can be found in Ref. \onlinecite{Kop/dwave}.

\section{Quasiparticle conductivity}\label{CONDUCTIVITY}

Consider the a.c. quasiparticle conductivity which can be observed if
vortices are pinned. The distribution function can be found from Eqs. (\ref
{kineq1}, \ref{kineq2}). We are looking again for the distribution function $%
{\sl f}_1$ which is independent of coordinates. One has 
\begin{eqnarray*}
e\left( {\bf v}_{F}\cdot {\bf E}\right) g_{-}\frac{\partial {\sl f}^{(0)}}{%
\partial \epsilon }+\left( {\bf v}_{F}\cdot \nabla \right) (g_{-}f_{2})+g_{-}%
\frac{\partial {\sl f}_{1}}{\partial t} &&  \nonumber \\
+\left[ \frac{e}{c}\left[ {\bf v}_{F}\times {\bf H}\right] g_{-}-\frac{1}{2}%
\left( f_{-}\hat{\nabla}\Delta _{{\bf p}}^{*}+f_{-}^{\dagger }\hat{\nabla}%
\Delta _{{\bf p}}\right) \right] \cdot \frac{\partial {\sl f}_{1}}{\partial 
{\bf p}} &=&J .
\end{eqnarray*}
We omit the time derivatives of $\Delta$ because vortices do not move. After
averaging over the vortex lattice we get 
\begin{eqnarray*}
\pi \left[ {\bf z}\times {\bf v}_{\perp }\right] \cdot \frac{\partial {\sl f}%
_{1}}{\partial {\bf p}}-\frac{\partial {\sl f}_{1}}{\partial t}%
\int_{S_{0}}g_{-}\,d^{2}r+\int_{S_{0}}J\,d^{2}r \\
=e\left( {\bf v}_{F}\cdot {\bf E}\right) \frac{\partial {\sl f}^{(0)}}{%
\partial \epsilon }\int_{S_{0}}g_{-}\,d^{2}r \, .
\end{eqnarray*}

For the distribution function in the form 
\begin{equation}
{\sl f}_{1}=-\frac{\partial {\sl f}^{(0)}}{\partial \epsilon } [({\bf E\cdot
p}_{\perp })\tilde{\gamma}_{{\rm O}}-([{\bf E\times p}_{\perp }]\cdot {\bf 
\hat{z}})\tilde{\gamma}_{{\rm H}}]
\end{equation}
we obtain 
\begin{eqnarray}
\frac{\partial \tilde{\gamma}_{{\rm O}}}{\partial \alpha }-\tilde{\gamma}_{%
{\rm H}}-V\left( \alpha \right) \tilde{\gamma}_{{\rm O}} &=&-\frac{e}{%
m\omega _{c}}g_{-} \, , \nonumber \\
\frac{\partial \tilde{\gamma}_{{\rm H}}}{\partial \alpha }+\tilde{\gamma}_{%
{\rm O}}-V\left( \alpha \right) \tilde{\gamma}_{{\rm H}} &=&0 \, .
\end{eqnarray}

The solution is 
\begin{equation}
W_{\pm }=\left[ \mp \frac{ie}{m\omega _{c}}\int_{0}^{\alpha }g_{-}e^{\mp
i\alpha ^{\prime }-F\left( \alpha ^{\prime }\right) }d\alpha ^{\prime
}+C_{\pm }\right] e^{\pm i\alpha +F\left( \alpha \right) } . \label{Wcond}
\end{equation}
The periodicity condition $W\left( 0\right) =W\left( \pi /2\right) $ gives
for $\epsilon \ll \Delta _{0}$ 
\begin{equation}
C_{\pm } =\frac{e}{m}\frac{1}{\left[ -i\omega +\left\langle
g_{-}\right\rangle /\tau \right] }\left[ \frac{\sinh \lambda }{\cosh
2\lambda }\pm i\left( 1-\frac{\cosh \lambda }{\cosh 2\lambda }\right) \right]
\label{Ccond}
\end{equation}
where $\lambda =F(\alpha _\epsilon )$.

The quasiparticle current is 
\begin{eqnarray*}
{\bf j}^{\left( qp\right) } &=&-\nu \left( 0\right) e\int {\bf v}_{F}g_{-}%
{\sl f}_{1}d\epsilon \frac{d\Omega }{4\pi } \\
&=&\sigma _{{\rm O}}^{(qp)} {\bf E}+\sigma _{{\rm H}}^{(qp)} [{\bf E} \times 
{\bf z}]
\end{eqnarray*}
where 
\begin{equation}
\sigma _{{\rm O,H}}^{(qp)}= \frac{\nu \left( 0\right) e}{2}\int v_{\perp
}p_{\perp }g_{-}\tilde{\gamma}_{{\rm O,H}}\frac{d\Omega }{4\pi }\frac{%
\partial {\sl f}^{\left( 0\right) }}{\partial \epsilon }d\epsilon
\, . \label{sigmas}
\end{equation}
Calculating the integral over $d\alpha$ in Eq. (\ref{sigmas})  we find 
\begin{eqnarray}
\left\langle \tilde{\gamma}_{{\rm O}}g_{-}\right\rangle _{\alpha } &=&\frac{%
\pi e}{4m\omega _{c}}\frac{\left\langle g_{-}\right\rangle ^{2}}{\lambda ^{2}%
}\left( \lambda -\frac{\tanh \lambda }{\tanh ^{2}\lambda +1}\right)
\, , \label{aver1} \\
\left\langle \tilde{\gamma}_{{\rm H}}g_{-}\right\rangle _{\alpha } &=&\frac{%
\pi e}{4m\omega _{c}}\frac{\left\langle g_{-}\right\rangle ^{2}}{\lambda ^{2}%
}\frac{\tanh ^{2}\lambda }{\tanh ^{2}\lambda +1} \, , \label{aver2}
\end{eqnarray}
and 
\[
\lambda =\frac{\pi }{4}\left\langle g_{-}\right\rangle \left[ -i\omega
+\left\langle g_{-}\right\rangle /\tau \right] \omega _{c}^{-1} \, .
\]
Since $\lambda $ is independent of the momentum directions, the
quasiparticle conductivity becomes 
\begin{equation}
\sigma _{{\rm O}}^{(qp)} =Ne\int \left\langle g_{-}\tilde{\gamma}_{{\rm O}%
}\right\rangle \frac{\partial {\sl f}^{(0)}}{\partial \epsilon }\frac{%
d\epsilon }{2}
\end{equation}
and the same expression for $\sigma _{{\rm H}}^{(qp)}$ where $\left\langle
g_{-}\tilde{\gamma}_{{\rm O}}\right\rangle$ is replaced with $\left\langle
g_{-}\tilde{\gamma}_{{\rm H}}\right\rangle$.

Consider first the superclean limit $\omega _{c}\tau _{eff}\gg T/T_{c}$ such
that ${\rm Re}\lambda \ll 1$. For $\omega \tau _{eff}\gg 1$, where $\tau
_{eff}\sim \left( T_{c}/T\right) \tau $ the response Eq. (\ref{aver1}) has
resonances at $i\lambda =(2M+1)\pi /4$ which is again the condition of Eq. (%
\ref{resonance}): 
\begin{eqnarray*}
\left\langle g_{-}\tilde{\gamma}_{{\rm O}}\right\rangle &=&\frac{4e}{\pi
m\omega _{c}}\sum_{M}\frac{\left\langle g_{-}\right\rangle
^{2}E_{0}(\epsilon )}{(2M+1)^{2}}\delta \left[ \omega -(2M+1)E_{0}(\epsilon
)\right] \\
&& +\frac{ie\left\langle g_{-}\right\rangle }{m\omega } \, .
\end{eqnarray*}
The dissipative part of the quasiparticle conductivity becomes 
\[
{\rm Re}\, \sigma _{{\rm O}}^{(qp)}= \frac{2Ne^{2}}{\pi mT}\frac{\omega
_{c}^{2}\Delta _{0}}{\left| \omega \right| ^{3}}\sum_{M=0}^{\infty }\cosh
^{-2}\left[ \frac{\Delta _{0}\omega _{c}(2M+1)}{2T\left| \omega \right| }%
\right] .
\]
It is 
\[
{\rm Re}\, \sigma _{{\rm O}}^{(qp)}=\frac{2Ne^{2}\omega _{c}}{\pi m\omega
^{2}} 
\]
for $\omega \gg E_{g}$ where $E_{g}=\Delta _{0}\omega _{c}/2T$, but
decreases exponentially for smaller $\omega \ll E_{g}$. The dissipative part
for $\omega \ll E_{g}$ is mostly due to $\tau $. Since $\lambda \ll 1$ one
has in this limit 
\[
{\rm Re}\, \sigma _{{\rm O}}^{(qp)} =\frac{\pi Ne^{2}}{3m\omega _{c}} \int 
\frac{\partial {\sl f}^{(0)}}{\partial \epsilon }\frac{d\epsilon }{2}%
\left\langle g_{-}\right\rangle ^{2}\lambda =\frac{5\pi ^7T^4}{24 m\omega
_c^2\tau \Delta _0^4}\, . 
\]

On the moderately clean side, such that $\omega _{c}\tau _{eff}\ll T/T_{c}$
one has ${\rm Re}\lambda \gg 1$. The conductivity has a Drude form 
\[
\sigma _{{\rm O}}^{(qp)}=\frac{Ne^{2}}{m}\int \frac{\partial {\sl f}^{(0)}}{%
\partial \epsilon }\frac{d\epsilon }{2}\frac{\left\langle g_{-}\right\rangle 
}{\left[ -i\omega +\left\langle g_{-}\right\rangle /\tau \right] }\, .
\]

The Hall conductivity does not contain contributions from poles because the
resonances in Eq. (\ref{aver2}) with $M>0$ cancel those with $M<0$. For $%
\omega \ll \left\langle g_{-}\right\rangle /\tau $ one has 
\[
\left\langle \tilde{\gamma}_{{\rm H}}g_{-}\right\rangle _{\alpha }=\frac{%
e\tau }{m}\frac{\tanh ^{2}w}{(\tanh ^{2}w+1)w} 
\]
where $w\left( \epsilon \right) =\pi \left\langle g_{-}\right\rangle
^{2}/4\omega _{c}\tau $. The conductivity is 
\[
\sigma _{{\rm H}}^{(qp)}=\frac{Ne^{2}\tau }{m}\int_{0}^{\infty }\frac{%
\partial {\sl f}^{(0)}}{\partial \epsilon }\frac{\tanh ^{2}w}{\tanh ^{2}w+1}%
\frac{d\epsilon }{w} \, .
\]

If $T\ll \Delta _{0}\sqrt{\omega _{c}\tau }$ one has $\epsilon \sim T$ and $%
w\ll 1$. In this limit 
\[
\sigma _{{\rm H}}^{(qp)}=\frac{\pi Ne^{2}}{4m\omega _{c}}\int_{0}^{\infty }%
\frac{\partial {\sl f}^{(0)}}{\partial \epsilon }\left\langle
g_{-}\right\rangle ^{2}d\epsilon =\frac{\pi ^{3}Ne^{2}T^{2}}{12m\omega
_{c}\Delta _{0}^{2}} \, .
\]
If $T\gg \Delta _{0}\sqrt{\omega _{c}\tau }$ and $\omega _{c}\tau \ll 1$ the
integral is determined by $w\sim 1$ and $\epsilon \sim \Delta _{0}\sqrt{%
\omega _{c}\tau }\ll T$. We have 
\[
\sigma _{{\rm H}}^{(qp)} =\frac{0.39Ne^{2}\tau \Delta _{0}\sqrt{\omega
_{c}\tau }}{Tm} \, .
\]

\section{Conclusions}

We discuss and analyze the ``Landau levels'' {\it vs} ``energy bands'' 
opposition concerning the structure of the excitation spectrum in the 
mixed state of
superconductors, and in particular, of $d$--wave superconductors. We find
that the actual picture of quantization is an interplay  between the two 
limiting  images of the energy spectrum. Our
analysis shows that the influence of the magnetic field on delocalized
excitations in a superconductor can not be reduced to a mere action of the
effective vortex lattice potential. In fact, magnetic field has a two-fold
effect: On one hand, it creates vortices and thus provides a periodic
potential for excitations, on the other hand, it also affects a long range
motion of quasiparticles in a manner similar to that in normal metals. For
low energy excitations, the long range effects are less pronounced. However,
excitations with energies $\epsilon >\Delta _{0}\sqrt{H/H_{c2}}$ mostly 
show the long range quantization. The energy spectrum consists of ``Landau
levels'' which are split into bands by the periodic vortex potential. In the
quasiclassical approximation $p_{F}\xi \gg 1$, the bandwidth is of the order
of the distance between the Landau levels; it is small compared to the
energy itself.

An a.c. electric field induces transitions between the states belonging
to different Landau levels. Using the microscopic kinetic equations we
demonstrate that these transitions can be seen as an increase in the vortex
friction and/or in the quasiparticle conductivity due to a resonant
absorption at frequencies corresponding to the energy differences between
the Landau levels.

\acknowledgements
We are grateful to G. Blatter, A. Mel'nikov, Z. Te\v {s}anovi\'{c}, and G.
Volovik for illuminating discussions. This research is supported by
US DOE, grant W-31-109-ENG-38,  and by NSF, STCS \#DMR91-20000 and by Russian
Foundation for Basic Research grant 99-02-16043.

\end{multicols}
\end{document}